\documentclass{aa}

\usepackage{graphicx}
\usepackage{txfonts}
\usepackage{longtable}
\usepackage{lscape}
\usepackage{natbib}
\usepackage[font=small]{caption}

\begin{document}

\title{On the physical origin of the second solar spectrum \\
of the Sc~{\sc ii} line at 4247~{\AA}}

\author{Luca Belluzzi}

\institute{Dipartimento di Astronomia e Scienza dello Spazio, University of 
Firenze, Largo E. Fermi 2, I-50125 Firenze, Italy}



\abstract
{The peculiar three-peak structure of the linear polarization profile shown in 
the second solar spectrum by the Ba~{\sc ii} line at 4554~{\AA} has been 
interpreted as the result of the different contributions coming from the barium 
isotopes with and without hyperfine structure.
In the same spectrum, a triple peak polarization signal is also observed in 
the Sc~{\sc ii} line at 4247~{\AA}. Scandium has a single stable isotope 
($^{45}$Sc), which shows hyperfine structure due to a nuclear spin $I=7/2$.}
{We investigate the possibility of interpreting the linear polarization profile 
shown in the second solar spectrum by this Sc~{\sc ii} line in terms of 
hyperfine structure.}
{A two-level model atom with hyperfine structure is assumed.
Adopting an optically thin slab model, the role of atomic polarization and of 
hyperfine structure is investigated, avoiding the complications caused by
radiative transfer effects. 
The slab is assumed to be illuminated from below by the photospheric 
continuum, and the polarization of the radiation scattered at 90$^{\circ}$ 
is investigated.}
{The three-peak structure of the scattering polarization profile observed in 
this Sc~{\sc ii} line cannot be fully explained in terms of hyperfine 
structure.}
{Given the similarities between the Sc~{\sc ii} line at 4247~{\AA} and the 
Ba~{\sc ii} line at 4554~{\AA}, it is not clear why, within the same modeling 
assumptions, only the three-peak $Q/I$ profile of the barium line can be fully 
interpreted in terms of hyperfine structure.
The failure to interpret this Sc~{\sc ii} polarization signal raises 
important questions, whose resolution might lead to significant improvements 
in our understanding of the second solar spectrum. 
In particular, if the three-peak structure of the Sc~{\sc ii} signal is 
actually produced by a physical mechanism neglected within the approach 
considered here, it will be extremely interesting not only to identify this 
mechanism, but also to understand why it seems to be less important in the 
case of the barium line.}


\maketitle

\section{Introduction}
\label{sect:intro}
The theoretical interpretation of the ``second solar spectrum'', namely the 
linearly polarized spectrum of the solar radiation coming from quiet 
regions close to the limb, is presently one of the most intriguing challenges 
in the field of solar physics. 
Although the basic physical process at the origin of this spectrum is clear
(scattering line polarization), and although several of its properties and 
peculiarities have been interpreted through the theoretical approaches 
that have been proposed so far, our understanding of this spectrum 
remains rather fragmentary, and several features still elude any attempt 
of interpretation.
The main difficulty in the interpretation of the second solar spectrum is that 
many physical mechanisms are capable of generating or modifying the 
polarization of the solar radiation, and it is an extremely complicated task 
to properly quantify their effects in such a complex environment as the solar 
atmosphere.
On the other hand, our knowledge of some of these mechanisms is still rather 
poor, since they have received attention only recently, both from a theoretical 
and experimental point of view (e.g., evaluation of the depolarizing 
collisional rates, development of theoretical frameworks able to account 
for partial redistribution effects in a self-consistent way, etc.). 
Nevertheless, the efforts that have been made in this sense are fully 
justified, because a complete and correct understanding and modeling of the 
physics underlying the formation of the second solar spectrum will allow 
us to fully exploit its enormous diagnostic potential, mainly for the 
investigation of the magnetic fields present in the solar atmosphere 
\citep[see][for a recent review]{JTB09}.

\begin{figure*}[!t]
\centering
\includegraphics[width=\textwidth]{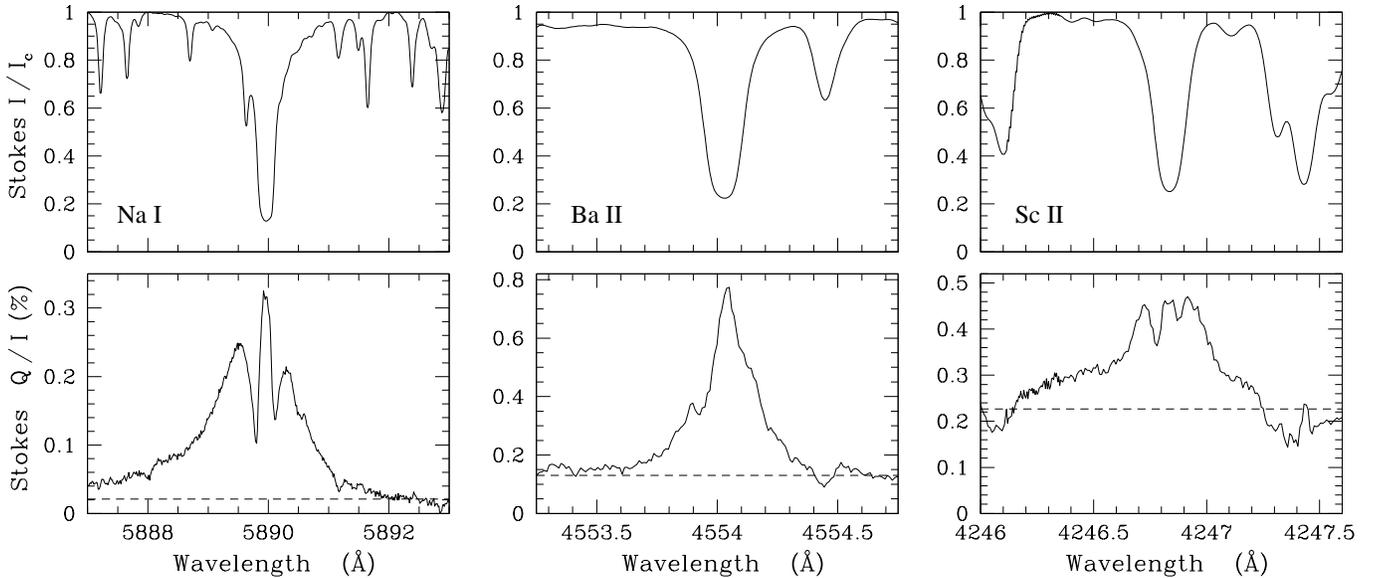}
\caption{Three spectral lines showing a three-peak $Q/I$ profile in the second 
solar spectrum: the Na~{\sc i} D$_2$ line at 5889~{\AA} (left), the Ba~{\sc ii}
line at 4554~{\AA} (middle), and the Sc~{\sc ii} line at 4247~{\AA} (right).
All the observations are taken from \citet{Gan00,Gan02}.}
\label{fig:3-peak}
\end{figure*}

Observations performed with instruments having sensitivities on the order of 
10$^{-3}$-10$^{-4}$ have shown in great detail the spectral richness and 
complexity of the second solar spectrum \citep[see][]{Ste96,Ste97}. 
Among the profiles observed, those with a three-peak structure have 
particularly excited the interest and curiosity of the scientific community.
Remarkable examples are the three-peak $Q/I$ profiles of the Ca~{\sc i} line 
at 4226~{\AA}, of the Na~{\sc i} D$_2$ line at 5889~{\AA}, and of 
the Ba~{\sc ii} D$_2$ line at 4554~{\AA}.
The intensity spectrum and the second solar spectrum of the Na~{\sc i} and 
Ba~{\sc ii} D$_2$ lines are shown in the first two panels of 
Fig.~\ref{fig:3-peak}.

The three-peak structure shown by the Ba~{\sc ii} D$_2$ line has been 
explained in terms of the presence of barium isotopes both with and without  
hyperfine structure (HFS) \citep[see][]{Ste97a,Bel07}. In particular, it has 
been shown that the two secondary peaks in the wings of the $Q/I$ profile are 
due to the isotopes with HFS ($\approx$ 18\% in abundance), while the central,
higher peak is produced by the isotopes without HFS ($\approx$ 82\% in 
abundance). Taking into account the effect of the HFS shown by this rather 
small fraction of barium isotopes, the observed three-peak profile could be 
reproduced to very high accuracy, even within the simplifying modeling 
assumption of the so-called optically thin slab model (i.e., neglecting 
radiative transfer effects).

As can be observed in the right panel of Fig.~\ref{fig:3-peak}, also the 
Sc~{\sc ii} line at 4247~{\AA} shows in the second solar spectrum a 
three-peak $Q/I$ profile. In contrast to the $Q/I$ profile produced by the 
Ba~{\sc ii} D$_2$ line, which shows a central peak (due to the isotopes 
without HFS) of amplitude much larger than that of the lateral peaks, 
the three peaks exhibited by this scandium signal have approximately the same 
amplitude. This peculiarity strongly suggests that also the 
three-peak structure of this signal might be due to HFS, since scandium has a
single stable isotope, which shows HFS (see Sect.~2.2).
This possibility is strengthened by the wavelength separation between the 
lateral peaks being very similar to that observed in the $Q/I$ 
profile of the Ba~{\sc ii} D$_2$ line, and by this scandium line in the 
intensity spectrum being very similar to the Ba~{\sc ii} D$_2$ line.
We note that these latter circumstances do not hold in the case of the 
Na~{\sc i} D$_2$ line: the sodium line is much stronger and broader in the 
intensity spectrum, and the wavelength separation between the lateral peaks 
shown by the $Q/I$ profile is much larger than for either barium or scandium. 
Indeed, it has already been observed that the three-peak structure of the 
Na~{\sc i} D$_2$ line cannot be explained only in terms of the HFS 
exhibited by the single stable isotope of sodium, but that its interpretation 
seems to require the inclusion of other physical ``ingredients'' such as 
``super-interferences'' and lower level polarization \citep[see][]{Lan98}, 
and/or the effects of partial redistribution in frequency \citep[see][]{Hol05}, 
and/or the enhancement of the line-center scattering polarization peak by 
vertical magnetic fields \citep[see][]{JTB02}.
For the reasons explained above, we considered it worthwhile to 
investigate whether the peculiar three-peak structure of this scandium $Q/I$ 
signal might be explained in terms of HFS, within a modeling approach similar 
to that proposed by \citet{Bel07} for the Ba~{\sc ii} D$_2$ line.

\section{Formulation of the problem}
\subsection{The two-level atom with hyperfine structure}
It is well-known that hyperfine structure is produced by the influence of the 
nucleus on the energy levels of the atom. 
On the one hand, this influence is related to the nuclei of the various 
isotopes having slightly different volumes and masses (isotopic effect), and 
on the other hand, to the coupling of the nuclear spin $\vec{I}$ with the total 
angular momentum $\vec{J}$ of the electronic cloud (nuclear spin effect).
We note that it is customary to speak about hyperfine structure {\it tout 
court} when referring to the nuclear spin effect only.

In the absence of magnetic fields, using Dirac's notation, the energy 
eigenvectors of an atomic system with HFS can be written in the form 
$|\, \alpha J I F f \!>$, where $\alpha$ represents a set of inner quantum 
numbers (specifying the electronic configuration and, if the atomic system is 
described by the $L$-$S$ coupling scheme, the total electronic orbital, and 
spin angular momenta), $F$ is the quantum number associated with the total 
angular momentum operator (electronic plus nuclear: $\vec{F}=\vec{J}+\vec{I}$), 
and $f$ is the quantum number associated with the projection of $\vec{F}$ along 
the quantization axis.
It is possible to demonstrate that the HFS Hamiltonian, describing the
interaction between the nuclear spin and the electronic angular momentum,
can be expressed as a series of electric and magnetic multipoles
\citep[see, for example,][]{Kop58}.
In this investigation, we retain only the first two terms (magnetic-dipole
and electric-quadrupole), which are given by
\begin{eqnarray}
	\label{eq:hfs-energy}
        < \alpha J I F f \, | \, H_{\rm hfs}^{(1)} \, |\, \alpha J I F^{\prime}
	f^{\prime} \! >  & = & \delta_{F F^{\prime}}
	\, \delta_{f f^{\prime}} \, \frac{\mathcal{A}(\alpha,J,I)}{2}K 
	\nonumber \;\; , \\
	< \alpha J I F f \, | \, H_{\rm hfs}^{(2)} \, |\, \alpha J I F^{\prime}
	f^{\prime} \! > & = & \delta_{F F^{\prime}}
	\, \delta_{f f^{\prime}} \,  \mathcal{B}(\alpha,J,I) \nonumber \\
	& & \!\!\!\!\!\!\!\!\!\!\!\!\!\!\!\!\!\!\!\!\!\! \times \frac{3}{8} 
	\Bigg\{ 
	\frac{K(K+1)-\frac{4}{3} J(J+1)I(I+1)}{I(2I-1)J(2J-1)} 
	\Bigg\} \;\; ,
\end{eqnarray}
where $\mathcal{A}(\alpha,J,I)$ and $\mathcal{B}(\alpha,J,I)$ are the
magnetic-dipole and the electric-quadrupole HFS constants, respectively, and 
where
\begin{equation}
        K=F(F+1)-J(J+1)-I(I+1) \;\; .
\end{equation}

We describe the atomic system by means of the density matrix formalism, a 
robust theoretical framework very suitable for handling atomic 
polarization phenomena (population unbalances and quantum interferences
between pairs of magnetic sublevels) that can be induced in the atomic system 
(for example by an anisotropic incident radiation field).
Since the upper and lower levels of the Sc~{\sc ii} line at 4247~{\AA} pertain 
to singlet terms (this is the only line of the multiplet), we consider 
a simple two-level model atom with HFS. Following \citet{LL04} (hereafter 
LL04), we describe the atom through the density matrix elements 
\begin{equation}
\label{eq:rhoFFp}
	< \alpha J I F f \, |\,\, \hat{\rho} \, | \, \alpha J I F^{\prime} 
	f^{\prime} \!> \, = \, \rho_{\alpha J I} \, (F f, F^{\prime} 
	f^{\prime}) \;\; ,
\end{equation}
where $\hat{\rho}$ is the density-matrix operator.
We recall that the diagonal elements represent the populations of the various 
magnetic sublevels, while the off-diagonal elements represent the quantum 
interferences, or coherences, between pairs of magnetic sublevels. 
In the following, we work in terms of the so-called spherical statistical 
tensors. 
The conversion of the density matrix elements of Eq.~(\ref{eq:rhoFFp}) into 
the spherical statistical tensor representation is given by the relation
\begin{eqnarray}
	^{\alpha J I}\rho^K_Q(F,F^{\prime}) & = & \sum_{ff^{\prime}} (-1)^{F-f} 
	\sqrt{2K+1} \nonumber \\
	& & \times \,
	\Bigg( \begin{array}{ccc}
		F & \!F^{\prime} & \!K \\
		f & \!-f^{\prime} & \!-Q
	\end{array} \Bigg) \;
	\rho_{\alpha J I} \, (F f, F^{\prime} f^{\prime}) \;\; .
\end{eqnarray}
The statistical equilibrium equations (SEE), and the radiative transfer 
coefficients for a two-level atom with HFS, in the spherical statistical 
tensor representation can be found in LL04. 
Here we write only the expression of the emission coefficient (in the absence 
of magnetic fields)
\begin{eqnarray}
	& & \varepsilon_{j}(\nu,\mathbf{\Omega})=
	\frac{h\nu}{4\pi}\mathcal{N} (2J_{u}+1)
	A(\alpha_{u} J_{u} \to \alpha_{\ell} J_{\ell}) \nonumber \\
	& & \times \sum_{KQ} \sum_{F_{u} F_{u}^{\prime} F_{\ell}}
	(-1)^{1+F_{\ell}+F_{u}^{\prime}} (2F_{\ell} +1)
	\sqrt{3(2F_{u}+1)(2F_{u}^{\prime}+1)} \nonumber \\
	& & \times \, \bigg\{ \begin{array}{ccc}
	       J_{u} & J_{\ell} & 1 \\
               F_{\ell} & F_{u} & I
	\end{array} \bigg\}
	\, \bigg\{ \begin{array}{ccc}
	       J_{u} & J_{\ell} & 1 \\
	       F_{\ell} & F_{u}^{\,\prime} & I
	\end{array} \bigg\}
	\, \bigg\{ \begin{array}{ccc}
	       1 & 1 & K \\
	       F_{u} & F_{u}^{\,\prime} & F_{\ell}
	\end{array} \bigg\} \nonumber \\
	& & \times \, {\mathcal{T}}_{Q}^{K}(j,\mathbf{\Omega})\,
	^{\alpha_{u}J_{u}I}\!\rho_{Q}^{K}(F_{u}^{\prime},F_{u}) \nonumber \\
	& & \times \, \frac{1}{2} \Big[
        \,\Phi(\nu_{\alpha_u J_u I F_u, \alpha_{\ell} J_{\ell} I F_{\ell}}-\nu) 
        + \Phi(\nu_{\alpha_u J_u I F_u^{\prime}, \alpha_{\ell} J_{\ell} I 
	F_{\ell}}-\nu)^{\ast} \Big] \; ,
\label{eq:epsilon}
\end{eqnarray}
with $j\!=\!0,1,2,3$ (corresponding respectively to the Stokes parameters
$I,Q,U,$ and $V$), $\mathcal{N}$ the number density of atoms,
$A(\alpha_{u}J_{u} \to \alpha_{\ell} J_{\ell})$ the Einstein coefficient
for spontaneous emission, ${\mathcal{T}}_{Q}^{K}(j,\mathbf{\Omega})$ a
geometrical tensor (see Sect.~5.11 of LL04), and $\Phi$ the profile of the line.
The indices $\ell$ and $u$ have the usual meaning of lower and upper (level), 
respectively.

It is important to recall that Eq.~(\ref{eq:epsilon}), and the SEE 
needed to find the spherical statistical tensors $\rho^K_Q(F_u,F_u^{\prime})$
are valid under the {\it{flat-spectrum approximation}}.
For a two-level atom with HFS, this approximation requires that the incident 
radiation field should be flat (i.e., independent of frequency) across a 
spectral interval $\Delta\nu$ larger than the frequency intervals among the 
HFS levels, and larger than the inverse lifetimes of the same levels. 
Given the small frequency separation between the various HFS levels (see 
Fig.~\ref{fig:hfs-comp}), this appears to be a good approximation for the 
Sc~{\sc ii} line under investigation.

\subsection{The atomic model}
\begin{figure}[!b]
\centering
\includegraphics[width=0.5\textwidth]{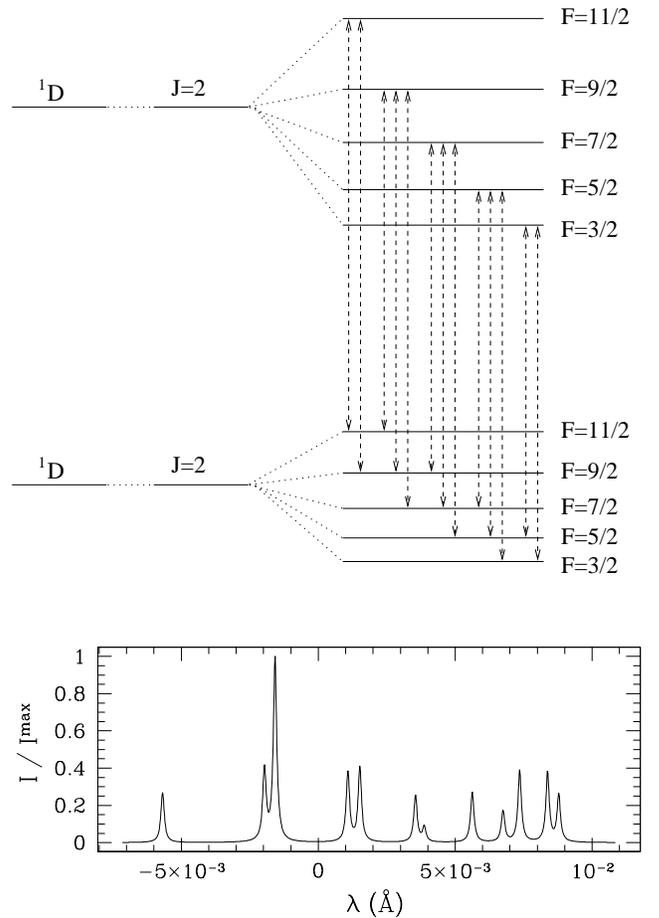}
\caption{Upper panel: Grotrian diagram showing the terms, the fine structure, 
and the hyperfine structure levels considered in the atomic model (the 
separation between the two $J$-levels is not on scale). 
The 13 HFS components of the line under investigation are drawn in the diagram. 
Lower panel: laboratory positions and relative strengths of the various HFS 
components, broadened by their natural width. 
Note that only twelve components are visible since two are blended. 
The zero of the wavelength scale is at 4246.82~{\AA}.}
\label{fig:hfs-comp}
\end{figure}

Scandium has only one stable isotope ($^{45}$Sc), of nuclear spin $I=7/2$.
The Sc~{\sc ii} line at 4247~{\AA} originates from the transition between the 
levels $3d4s\, ^1$D$_2$ (lower level) and $3d4p\, ^1$D$_2$ (upper level).
The Einstein coefficient for spontaneous emission is $A=1.29 \times 
10^{8}$~s$^{-1}$ \citep[][]{Ral08}.
Because of HFS, each $J$-level splits into 5 $F$-levels, and the spectral line 
under investigation consists of 13 HFS components (see the upper 
panel of Fig.~\ref{fig:hfs-comp}).
We use the energies of the $J$-levels provided by \citet{Ral08}, while 
we calculate the energies of the various HFS levels by applying 
Eq.~(\ref{eq:hfs-energy}), using the values of the magnetic dipole and of 
the electric quadrupole HFS constants listed in Table~\ref{Tab:const}.
\begin{table}[t]
\caption{Magnetic dipole and electric quadrupole HFS constants of the levels 
considered in this investigation. Data are taken from \citet{Vil92}.}
\centering
\begin{tabular}{ccccc}
	\hline
	\hline
	\noalign{\smallskip}
	Isotope & \multicolumn{2}{c}{$3d4s \, ^1$D$_2$} & 
	\multicolumn{2}{c}{$3d4p \, ^1$D$_2$} \\
	\noalign{\smallskip}
	 & $\mathcal{A}$~(MHz) & $\mathcal{B}$~(MHz) & 
	$\mathcal{A}$~(MHz) & $\mathcal{B}$~(MHz) \\
	\noalign{\smallskip}
        \hline
	\noalign{\smallskip}
	\noalign{\smallskip}
	$^{45}$Sc & 128.2 & -39 & 215.7 & 18 \\
	\noalign{\smallskip}
	\hline
\end{tabular}
\label{Tab:const}
\end{table}
The laboratory positions and the relative strengths of the various HFS 
components are shown in the lower panel of Fig.~\ref{fig:hfs-comp}.

Despite the simplicity of the atomic model considered (two-level atom), because 
of the high values of the quantum numbers involved, the SEE form a set of 3200 
equations in 3200 unknowns (the various spherical statistical tensors 
$^{\alpha J I}\rho^K_Q(Ff,F^{\prime}f^{\prime})$ of the upper and lower 
levels). To reduce the amount of numerical calculations involved in the 
solution of this set of equations, we limit ourselves to considering only the 
spherical statistical tensors with $K \le 2$, thus reducing the SEE to a set of 
278 equations. Because of the low value of the anisotropy factor that we assume 
for the radiation field (see following subsection), the statistical tensors of 
higher rank are found to be almost two orders of magnitude smaller, so that the 
results obtained within this approximation do not show any appreciable 
difference with respect to those obtained by carrying out the complete 
calculations.

\subsection{The optically thin slab model}
\begin{figure}[!t]
\centering
\includegraphics[width=0.4\textwidth]{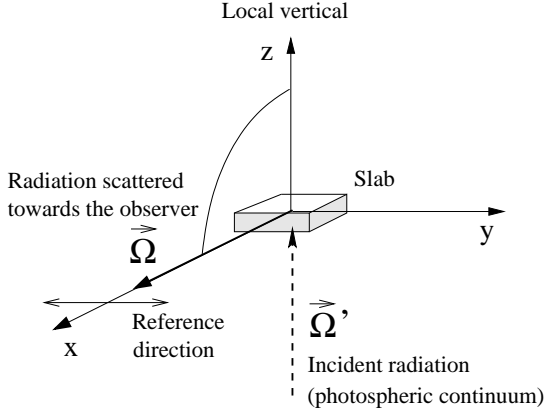}
\caption{Geometry of the problem being investigated.}
\label{fig:slab}
\end{figure}
To emphasize the atomic aspects involved in the problem, avoiding 
complications due to radiative transfer effects, we consider an optically thin 
slab of Sc~{\sc ii} ions located 1000~km above the solar surface (as defined in 
Sect.~12.3 of LL04), and we assume that the slab is illuminated from below by 
the photospheric continuum (see Fig.~\ref{fig:slab}).
Under these hypotheses, the atomic polarization can be calculated by 
solving the SEE directly for the given incident continuum radiation field. 
We remark that this model is basically the same as that employed 
by \citet{Bel07} to model the $Q/I$ profile of the Ba~{\sc ii} D$_2$ line.
We assume that the continuum is unpolarized and cylindrically symmetric 
about the local vertical. 
Under these assumptions, taking a reference system 
with the $z$-axis (the quantization axis) directed along the vertical, it can 
be shown that only two components of the radiation field tensor, in terms of 
which we describe the incident continuum, are non-vanishing
\begin{equation}
J^{0}_{0}(\nu) \! = \! \oint\frac{{\rm d}\Omega}{4\pi}I(\nu,\mu)
\;\; {\rm and} \;\;
J^{2}_{0}(\nu) \! = \! \frac{1}{2\sqrt{2}} \oint\frac{{\rm d}\Omega}{4\pi}
(3\mu^2-1)I(\nu,\mu) \, ,
\label{eq:JKQ}
\end{equation}
where $\mu$ is the cosine of the heliocentric angle.
The former quantity is the mean intensity of the radiation field (averaged over 
all directions), the second quantifies the anisotropy of the radiation 
field (imbalance between vertical and horizontal illumination). 
The mean intensity of the radiation field can also be expressed in terms of 
the average number of photons per mode, $\bar{n}$, while the anisotropy degree 
is often quantified through the so-called anisotropy factor, $w$. 
These new, non-dimensional quantities are related to $J^0_0$ and $J^2_0$ by 
the equations
\begin{equation}
	\bar{n}(\nu)=\frac{c^2}{2 h \nu^3} \, J^0_0(\nu) \;\; , \;\;\; 
	w(\nu)=\sqrt{2} \, \frac{J^2_0(\nu)}{J^0_0(\nu)} \;\; .
\end{equation}
We calculate the values of $\bar{n}(\nu)$ and $w(\nu)$ of the photospheric 
continuum following Sect.~12.3 of LL04 taking $h=1000$~km, and using the values 
of the disk center intensities, and of the limb-darkening coefficients given 
by \citet{Pie00}.
At the height of the slab, and at the frequency of the spectral line 
under investigation (4247~{\AA}) we find $\bar{n}=0.158 \times 10^{-2}$, 
and $w=0.189$.

Taking into account the atomic weight of scandium, assuming a temperature
of 6000~K, and neglecting microturbulent velocity, we obtain for this line a 
Doppler width, $\Delta \lambda_D$, given by
\begin{equation}
	\Delta \lambda_D= \frac{\lambda_0}{c}w_T = \frac{\lambda_0}{c}
	\sqrt{\frac{2k_BT}{M}}=21~{\rm m\AA} \;\; ,
\end{equation}
where $w_T$ is the thermal velocity, $T$ the kinetic temperature, $M$ the mass 
of the ion, and $k_B$ the Boltzmann constant.

\subsection{Polarization of the emergent spectral line radiation}
\begin{figure}
\centering
\includegraphics[width=0.5\textwidth]{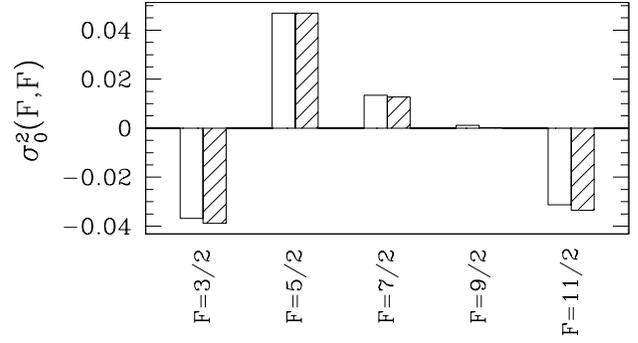}
\caption{Fractional atomic alignment 
$\sigma^2_0(F,F)=\rho^2_0(F,F)/\rho^0_0(F,F)$ in the lower (white columns) and 
upper (shaded columns) levels. Note that as far as the fractional atomic 
alignment of the lower level is concerned, the quantity 
$-\sigma^2_0(F_{\ell},F_{\ell})$ is plotted. The values have been obtained by
solving the SEE for the values of $\bar{n}$ and $w$ quoted in the text, and in 
the absence of collisions, magnetic fields, and stimulation effects.}
\label{fig:atomic-pol}
\end{figure}
Once the SEE have been written down and solved numerically, we can calculate 
the radiative transfer coefficients according to the equations of Sect.~7.9 of 
LL04.
We consider the radiation scattered by the slab at $90^{\circ}$, and we take 
the reference direction for positive $Q$ parallel to the slab (see 
Fig.~\ref{fig:slab}).
In the case of a tangential observation, under the approximation of a 
weakly polarizing atmosphere ($\varepsilon_I \gg \varepsilon_Q, \varepsilon_U, 
\varepsilon_V; \eta_I \gg \eta_Q, \eta_U, \eta_V, \rho_Q, \rho_U, \rho_V$), 
it can be shown that the emergent fractional polarization is given by
\citep[see][]{JTB03}
\begin{equation}
\label{eq:dichroism}
\frac{X(\nu,\mathbf{\Omega})}{I(\nu,\mathbf{\Omega})} \approx
\frac{\varepsilon_{X}(\nu,\mathbf{\Omega})}
{\varepsilon_{I}(\nu,\mathbf{\Omega})}-
\frac{\eta_{X}(\nu,\mathbf{\Omega})}
{\eta_{I}(\nu,\mathbf{\Omega})} \;\;\;\;\;\; {\rm{with}}\; X=Q,U,V \;\; .
\end{equation}
The first and second terms in the right-hand side of Eq.~(\ref{eq:dichroism}) 
represent the contribution to the emergent radiation due to
processes of selective emission and selective absorption (dichroism) of 
polarization components, respectively.

We note that while the lower level of Ba~{\sc ii} D$_2$, with $J=1/2$, can 
carry alignment only because of the presence of HFS, the lower level of this 
scandium line, with $J=2$, can be polarized 
also neglecting HFS. As a consequence, while in the case of the Ba~{\sc ii} 
D$_2$ line the lower level results to be significantly less polarized than the 
upper level \citep[see][]{Bel07}, in this scandium line the upper and lower 
levels carry the same amount of atomic polarization\footnote{Collisions will be neglected throughout this investigation.} (see Fig.~\ref{fig:atomic-pol}).
Although the contribution of dichroism is thus expected to be more important 
in this line than in the Ba~{\sc ii} D$_2$ line, we start our 
investigation by neglecting it. 
\begin{figure*}[!t]
\centering
\includegraphics[width=\textwidth]{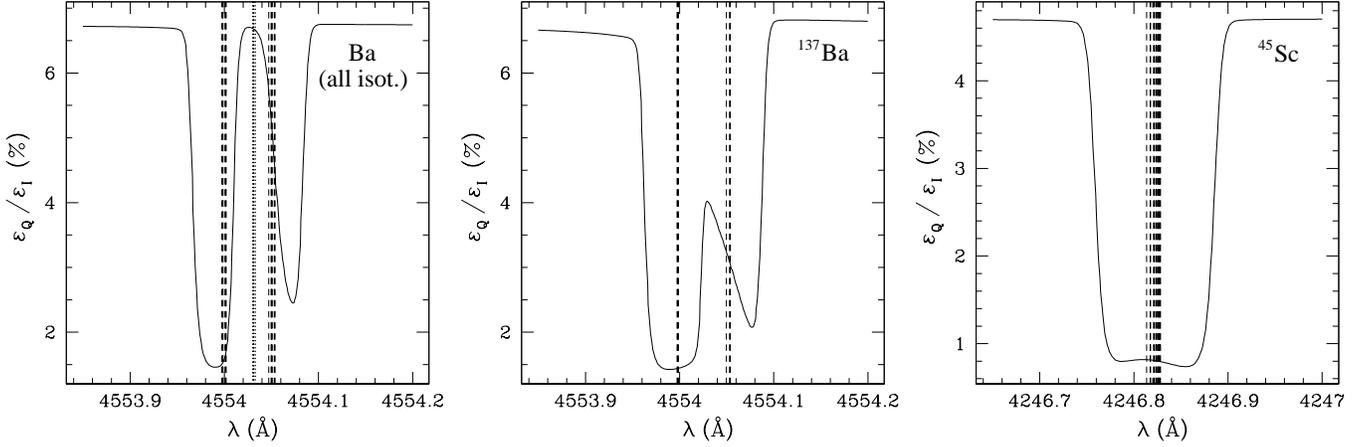}
\caption{Left panel: fractional polarization profile for the Ba~{\sc ii} line 
at 4554~{\AA}, obtained by applying Eq.~(\ref{eq:QsuI}) (without continuum).
The values of $\bar{n}$, $w$, and $\Delta \lambda_D$ are 
calculated for a slab of barium ions at 6000~K, located at a height of 
1000~km ($\bar{n}=0.225 \times 10^{-2}$, $w=0.176$, $\Delta \lambda_D=
13~{\rm m\AA}$). The dashed lines show the wavelength position of the HFS 
components of $^{135}$Ba and $^{137}$Ba, and the dotted lines the wavelength 
position of the fine structure components of the isotopes without HFS.
Middle panel: same as left panel but assuming that only the isotope $^{137}$Ba 
(with HFS) is present (100\% in abundance). 
The dashed lines show the wavelength position of the various HFS components.
Right panel: fractional polarization profile for the Sc~{\sc ii} line 
at 4247~{\AA}, obtained by applying Eq.~(\ref{eq:QsuI}) and assuming for 
$\bar{n}$, $w$, and $\Delta \lambda_D$ the values quoted in Sect.~2.3 
(corresponding to a slab at 6000~K, located 1000~km above the solar surface).
The dashed lines show the wavelength position of the various HFS components.}
\label{fig:frac-pol}
\end{figure*}
We therefore calculate the polarization of the radiation emerging from the slab 
by means of the equation
\begin{equation}
\label{eq:QsuI}
\frac{X(\nu,\mathbf{\Omega})}{I(\nu,\mathbf{\Omega})} \approx
\frac{\varepsilon_{X}(\nu,\mathbf{\Omega})}
{\varepsilon_{I}(\nu,\mathbf{\Omega})} \;\; .
\end{equation}
Some results obtained taking into account dichroism effects (i.e., applying 
Eq.~(\ref{eq:dichroism})) will be shown in Sect.~3.4.

We recall that the emission coefficient given in Eq.~(\ref{eq:epsilon}) 
includes only line processes. To reproduce, albeit qualitatively, 
the observed profile, we need to add the contribution of the 
continuum. Assuming that the continuum is constant across the line, we have
\begin{equation}
\label{eq:QsuI_cont}
\frac{X(\nu,\mathbf{\Omega})}{I(\nu,\mathbf{\Omega})} \approx
\frac{\varepsilon_{X}^{\,\ell}(\nu,\mathbf{\Omega})+\varepsilon_X^{\,c}}
{\varepsilon_{I}^{\,\ell}(\nu,\mathbf{\Omega})+\varepsilon_{I}^{\,c}} \;\; ,
\end{equation}
where the superscripts ``$c$'' and ``$\ell$'' denote that the corresponding 
quantities refer to continuum and line processes, respectively.
The quantities $\varepsilon_{X}^{\, c}$ and $\varepsilon_I^{\, c}$ are 
considered to be free parameters in the problem that are adjusted to
reproduce the observed polarization profile.

We point out that the three-peak structure of the $Q/I$ profile shown 
by the Ba~{\sc ii} D$_2$ line could be reproduced by \citet{Bel07} within 
the constraints of the same modeling assumptions and approximations 
outlined above.
However, we note that this ionized scandium line is a rather strong spectral 
line: as for Ba~{\sc ii} D$_2$, the wings of this line originate in the 
photosphere, while the line core originates in the high photosphere/low 
chromosphere. 
The optically thin slab model considered in this paper is therefore 
just a first order approximation. Nevertheless, as previously stated, it 
allows us to take into account in a very rigorous way the relevant atomic 
aspects of the problem, avoiding the complications produced by radiative 
transfer effects.

\section{Results}
\subsection{Three-peak structure of the Ba~{\sc ii} D$_2$ line $Q/I$ profile}
As discussed in \citet{Bel07}, the presence of barium isotopes both with and 
without HFS is understood to be important in explaining the three-peak 
structure of the $Q/I$ profile shown in the second solar spectrum by the 
Ba~{\sc ii} D$_2$ line. 
In particular, the central peak is due to the isotopes without HFS, 
while the two lateral peaks seem to be caused by the depolarizing effect of the 
isotopes with HFS ($^{135}$Ba and $^{137}$Ba), combined with the effect of the 
continuum, which depolarizes in the wings of the line.
The fractional polarization profile plotted in the left panel 
of Fig.~\ref{fig:frac-pol}, clearly shows the depolarizing effect of the 
isotopes with HFS, as well as the polarization enhancement at line center due 
to the isotopes without HFS.
This profile was obtained by applying Eq.~(\ref{eq:QsuI}) (i.e., neglecting 
the continuum) to the case of the Ba~{\sc ii} D$_2$ line, assuming for 
$\bar{n}$, $w$, and $\Delta \lambda_D$ the values corresponding to a slab of 
barium ions at 6000~K, located 1000~km above the solar surface. 
Starting from this fractional polarization profile, and adding the contribution 
of the continuum, \citet{Bel07} were able to obtain a good fit to the observed 
three-peak $Q/I$ profile.
\begin{figure*}[!t]
\centering
\includegraphics[width=\textwidth]{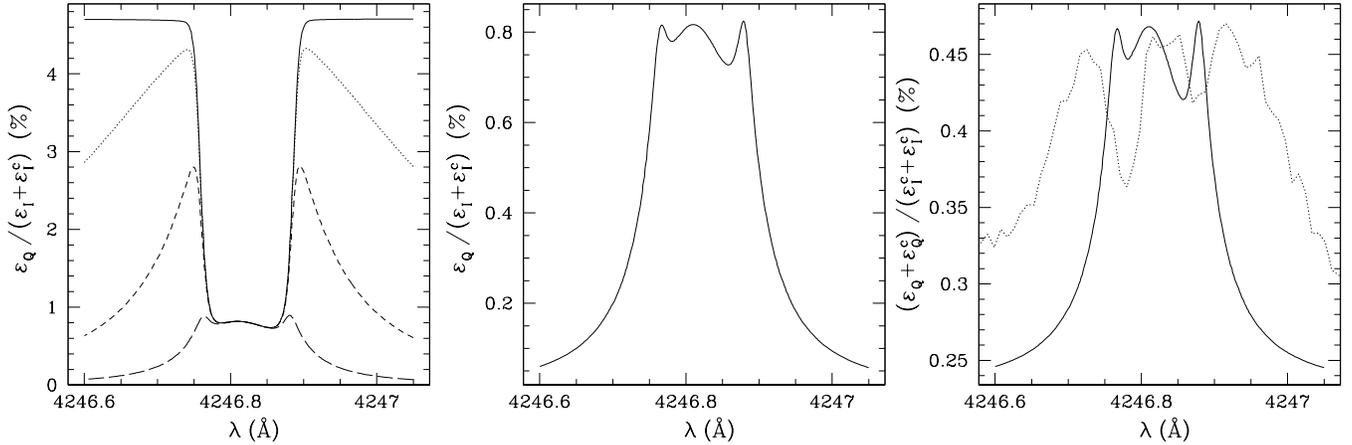}
\caption{Left panel: theoretical profiles obtained including the continuum 
contribution to the intensity. The various profiles have been obtained by 
assuming $\varepsilon_I^c/\varepsilon_I^{max}= 10^{-5}$ (dotted line), 
$10^{-4}$ (dashed line), and $10^{-3}$ (long-dashed line). 
The profile plotted with solid line has been obtained without continuum, and 
it is identical to the one shown in the right panel of Fig.~\ref{fig:frac-pol}. 
The other parameters ($\bar{n}$, $w$, and $\Delta \lambda_D$) have the same 
values as in the right panel of Fig.~\ref{fig:frac-pol}. 
Middle panel: profile obtained by setting $\varepsilon_I^c/\varepsilon_I^{max}= 
1.2 \times 10^{-3}$. 
Right panel: theoretical polarization profile obtained by applying 
Eq.~(\ref{eq:QsuI_cont}), setting $w=0.11$, 
$\varepsilon_I^c/\varepsilon_I^{max}= 2.4 \times 10^{-3}$,
$\varepsilon_Q^c/\varepsilon_I^c = 0.23 \times 10^{-2}$, and
$\Delta \lambda_D=21$~m{\AA} (solid line), and
the observed profile (dotted line) taken from \citet{Gan02}.} 
\label{fig:cont}
\end{figure*}

The fundamental role of the isotopes without HFS in producing the higher 
central peak of the observed linear polarization profile can be clearly 
appreciated from the middle panel of Fig.~\ref{fig:frac-pol}. 
Here the same fractional polarization profile as in the left panel of 
Fig.~\ref{fig:frac-pol} (without continuum) is plotted assuming that only the 
isotope $^{137}$Ba (with HFS) is present (100\% in abundance).
Comparing the two profiles, it is noticeable that although a polarization 
enhancement can be observed at line center also when only $^{137}$Ba is 
present, this is much larger when the isotopes without HFS are also taken 
into account. 
In particular, if these isotopes are considered, the polarization at line 
center reaches the same value as in the far wings.
This is an important peculiarity, since in this case there is no way, by adding 
the continuum, to obtain three peaks of the same amplitude (such as those 
observed in Sc~{\sc ii} line). 
On the other hand, this would be possible if, hypothetically, only the isotope 
$^{137}$Ba were present.

\subsection{Three-peak structure of the Sc~{\sc ii} 4247~{\AA} line $Q/I$ 
profile}
As previously pointed out, scandium has a single stable isotope, with
HFS, which is consistent with the interpretation of this scandium signal, 
showing three peaks of the same amplitude, in terms of HFS.
Applying Eq.~(\ref{eq:QsuI}), and using the values previously calculated for 
$\bar{n}$, $w$, and $\Delta \lambda_D$, we find the fractional polarization 
profile shown in the right panel of Fig.~\ref{fig:frac-pol}.
The profile clearly shows the depolarizing effect of HFS\footnote{Note that 
if HFS had been neglected, the fractional polarization profile calculated 
considering only the line processes (no continuum) would have been constant 
with wavelength, as it is clear from \citet{Bel07}, where the results obtained 
considering only $^{138}$Ba (without HFS) are shown.}, but the polarization 
enhancement at line-center, which produces the central peak, is 
by far less evident than in the case of $^{137}$Ba.
On the one hand, this is because the HFS components of $^{137}$Ba are gathered 
into two well separated groups (because of the large HFS splitting of the 
ground level), while the HFS components of scandium are gathered into a single 
group at line center, spreading out over a narrow wavelength interval of about 
15~m{\AA} (see the middle and right panels of Fig.~\ref{fig:frac-pol}, and 
Fig.~\ref{fig:hfs-comp}).
On the other hand, the smaller Doppler width assumed for barium (much heavier 
than scandium) contributes to making the line-core enhancement of the 
corresponding fractional polarization profile far more evident.

As in the case of barium, to reproduce the observed profile, it is 
necessary to include the effect of the continuum. 
By so doing, the value of the fractional polarization decreases in the 
wings of the line, and a two-peak profile with a small hump at line center is 
obtained. 
The left panel of Fig.~\ref{fig:cont} shows the profiles that result when 
assuming different values of $\varepsilon_I^c/\varepsilon_I^{max}$, where 
$\varepsilon_I^{max}$ is the maximum value of $\varepsilon_I^{\ell}$ in the 
wavelength range considered. A value of $\varepsilon_I^c/\varepsilon_I^{max}$ 
on the order of $10^{-3}$ is needed to obtain similar polarization 
values in the wing peaks and the line-core. 
Because of the small enhancement exhibited by the fractional polarization 
profile at line center, setting $\varepsilon_I^c/\varepsilon_I^{max}=1.2 \times 
10^{-3}$, we actually obtain a profile with three peaks of the same 
amplitude (see the middle panel of Fig.~\ref{fig:cont}).

To reproduce the amplitude of the central peak ($\approx$ 0.46\%, according to 
the observation of Gandorfer, 2002), and the polarization level of the 
continuum at the wavelength of this line ($\approx$ 0.23\%), we have to set 
$w=0.11$ and $\varepsilon_Q^c/\varepsilon_I^c = 0.23 \times 10^{-2}$. Assuming  
that $\varepsilon_I^c/\varepsilon_I^{max}= 2.4 \times 10^{-3}$ (which is a 
rather realistic value for a spectral line such as that under investigation), 
we obtain the theoretical profile plotted in the right panel of 
Fig.~\ref{fig:cont}. This profile indeed shows a three-peak structure, 
but its width can be immediately noticed to be much smaller than that of the 
observed profile. In particular, in the observed profile the wavelength 
separations between the two dips and between the two lateral peaks are  
about 90~m{\AA} and 190~m{\AA}, respectively, while in the theoretical profile
they are about 75~m{\AA} and 110~m{\AA}, respectively.
Moreover, while the amplitude of the dip at longer wavelengths (the 
``red'' dip) is very similar to the observed one, the amplitude of the ``blue''
dip is much smaller. 

\begin{figure*}
\centering
\includegraphics[width=\textwidth]{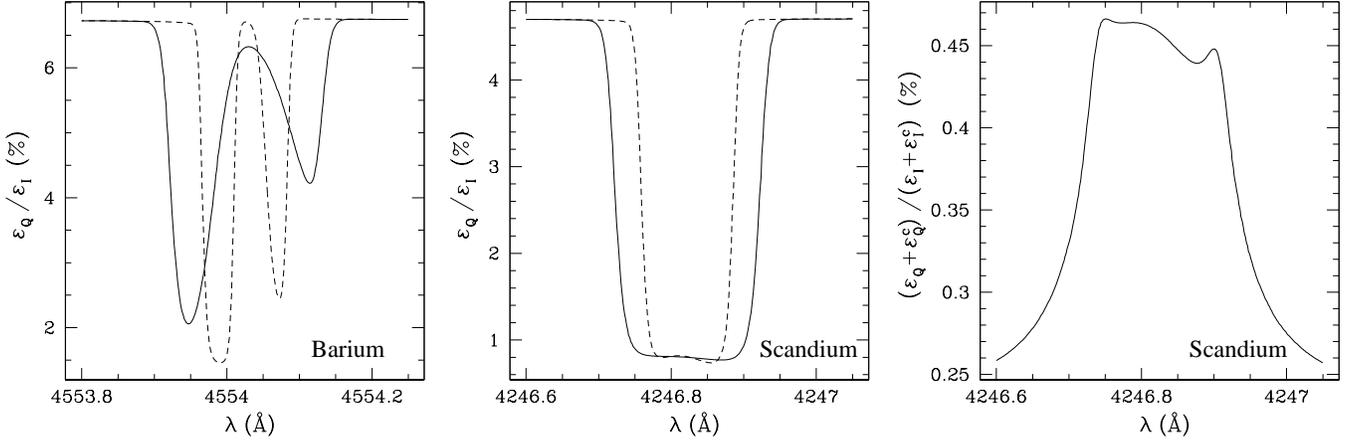}
\caption{Left panel: fractional polarization profiles for the Ba~{\sc ii} line 
at 4554~{\AA}, obtained by applying Eq.~(\ref{eq:QsuI}) (without continuum), 
assuming $\Delta \lambda_D=13$~m{\AA} (dotted line), and 
$\Delta \lambda_D=30$~m{\AA} (solid line). The other free parameters have the 
same values as in the left panel of Fig.~\ref{fig:frac-pol}. The dotted-line 
profile is the same as in the left panel of Fig.~\ref{fig:frac-pol}.
Middle panel: fractional polarization profiles for the Sc~{\sc ii} line at 
4247~{\AA}, obtained by applying Eq.~(\ref{eq:QsuI}), assuming 
$\Delta \lambda_D=21$~m{\AA} (dotted line), and $\Delta \lambda_D=33$~m{\AA} 
(solid line). The other free parameters have the same values as in the right 
panel of Fig.~\ref{fig:frac-pol}. The dotted-line profile is the same as in the 
right panel of Fig.~\ref{fig:frac-pol}.
Right panel: theoretical polarization profile for the Sc~{\sc ii} line, 
obtained by applying Eq.~(\ref{eq:QsuI_cont}) and setting
$\Delta \lambda_D=33$~m{\AA}, $w=0.11$, 
$\varepsilon_I^c/\varepsilon_I^{max}= 2.1 \times 10^{-3}$, and  
$\varepsilon_Q^c/\varepsilon_I^c = 0.23 \times 10^{-2}$.} 
\label{fig:doppler}
\end{figure*}

The only way to increase the width of the theoretical profile is to assume a 
larger value of the Doppler width. 
In the case of barium, to reproduce the observed profile it was 
necessary to assume a Doppler width of 30~m{\AA}, corresponding to a 
temperature of 6000~K and a microturbulent velocity of 1.8~km/s 
\citep[see][]{Bel07}.
In the left panel of Fig.~\ref{fig:doppler}, the barium fractional polarization 
profiles, obtained by assuming $\Delta \lambda_D=13$~m{\AA} (dotted line) and 
$\Delta \lambda_D=30$~m{\AA} (solid line) are shown: increasing the Doppler 
width, the wavelength separation between the two dips is increased, but the 
two-dip structure of the profile is not lost. 
In the middle panel of Fig.~\ref{fig:doppler}, we plot the fractional 
polarization profiles of the scandium line, obtained by assuming 
$\Delta \lambda_D=21$~m{\AA} (dotted line) and $\Delta \lambda_D=33$~m{\AA} 
(solid line), the latter corresponding to a temperature of 6000~K and a 
microturbulent velocity of 1.8~km/s. 
Increasing the Doppler width, the fractional polarization profile shows a 
significantly larger depolarizing feature, but the small polarization 
enhancement at line core, and therefore the three-peak structure of the 
polarization profile obtained by adding the continuum, are almost completely 
lost (see the middle and right panels of Fig.~\ref{fig:doppler}). 

\subsection{The effect of a unimodal microturbulent magnetic field}
\begin{figure*}[!t]
\centering
\includegraphics[width=\textwidth]{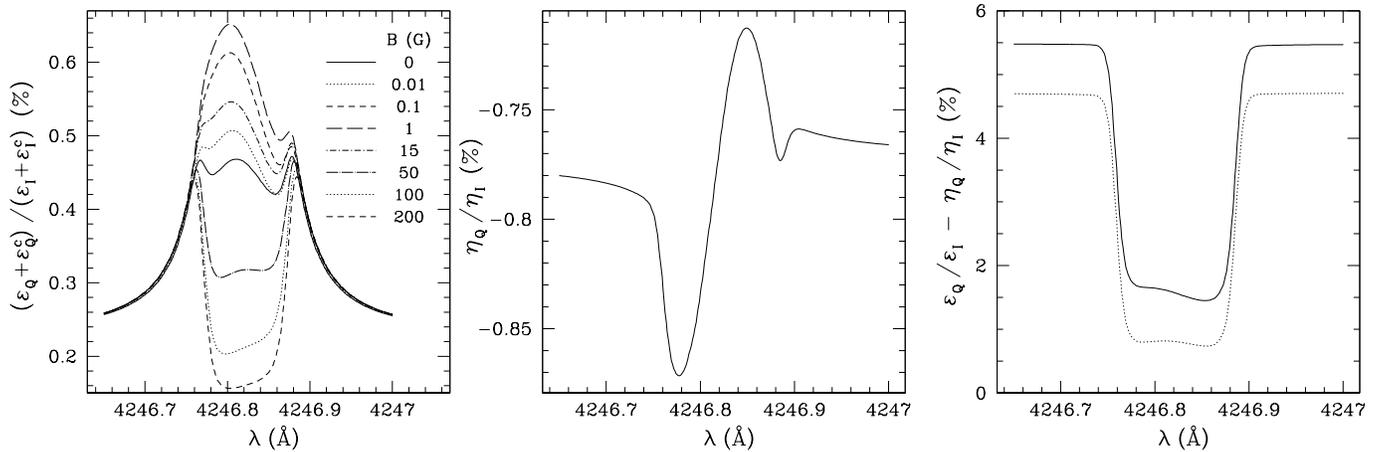}
\caption{Left panel: theoretical polarization profiles obtained in the 
presence of a unimodal microturbulent magnetic field of various intensities. 
All the profiles are calculated by using Eq.~(\ref{eq:QsuI_cont}), assuming for 
the free parameters the same values as in the right panel of 
Fig.~\ref{fig:cont}. The solid-line profile is the same as in the right 
panel of Fig.~\ref{fig:cont}. Middle panel: $\eta_Q/\eta_I$ profile 
calculated considering only the line processes (no continuum). The other 
free parameters have the same values as in the right panel of 
Fig.~\ref{fig:frac-pol}. Right panel: theoretical profiles obtained in the 
absence of continuum taking into account (solid line) and neglecting (dotted 
line) dichroism. The profiles were obtained by applying 
Eq.~(\ref{eq:dichroism}) and Eq.~(\ref{eq:QsuI}), respectively. 
The other free parameters have the same values as in the right panel of 
Fig.~\ref{fig:frac-pol}. The dotted-line profile is the same as in the right 
panel of Fig.~\ref{fig:frac-pol}.}
\label{fig:mag-dic}
\end{figure*}

The left panel of Fig.~\ref{fig:mag-dic} shows the theoretical linear 
polarization profiles obtained according to Eq.~(\ref{eq:QsuI_cont}), in the 
presence of a unimodal microturbulent magnetic field of various intensities.
A useful estimate of the spectral line sensitivity to the Hanle effect can be 
obtained by calculating the critical fields of the upper and lower 
levels \citep[see, for example,][]{JTB01}
\begin{equation}
	B_{\rm c}^{\rm up}=\frac{1.137 \times A}{g_u} \approx 15~{\rm G}
	\; , \;\;\; 
	B_{\rm c}^{\rm low}=\frac{1.137\times A\bar{n}}{g_{\ell}} \approx 20~
	{\rm mG} \; ,
\end{equation}
where the Einstein coefficient for spontaneous emission must be expressed in 
units of $10^7$ s$^{-1}$.
In agreement with the previous estimate, we observe an increase of the 
polarization, due to the lower-level Hanle effect, for magnetic fields between 
10~mG and 1~G. 
The polarization is found to increase by a larger amount at the wavelength 
positions of both the central peak and the blue dip, than of the red dip. 
If the continuum is changed, so that the wing peaks have the same amplitude 
as the central peak, a three-peak structure might possibly be recovered, but 
a red dip much deeper than the blue one would be found, in disagreement with 
the observation.
We point out that the lower-level Hanle effect is particularly evident in this 
scandium line because, as previously observed, the lower level carries the 
same amount of atomic polarization as the upper level (in the Ba~{\sc ii} 
D$_2$ line, where the lower level is much less polarized than the upper level, 
this effect cannot be observed).

If the magnetic field is further increased, the upper level Hanle effect 
becomes dominant, and a depolarization at the line core is observed.
We note that for magnetic fields on the order of 100~G, the polarization at 
line center becomes lower than in the far wings (i.e., lower than the 
continuum), although always remaining positive. 
A saturation regime is reached for fields of about 200~G.
Even though for fields on the order of 50~G, a three-peak structure can still 
be recovered by adjusting the parameters ($w$ and $\varepsilon_I^c$),  
nevertheless the observed profile cannot be reproduced.
In particular, the width of the theoretical profile does not 
increase appreciably with the magnetic field intensity, so that the 
disagreement with the observation previously discussed cannot be solved.

In conclusion, within our modeling assumptions, which also include the effect 
of a unimodal microturbulent magnetic field, it is not possible to reproduce 
the observed $Q/I$ profile.

\subsection{Results obtained including dichroism} 
In the middle panel of Fig.~\ref{fig:mag-dic}, the $\eta_Q/\eta_I$ profile is 
plotted. In the line-core, it exhibits an anti-symmetrical shape, while in 
the far wings, because of the presence of lower level polarization, it 
reaches an asymptotic value that differs from zero. 
The profile obtained by applying Eq.~(\ref{eq:dichroism}) is plotted with 
a solid line in the right panel of Fig.~\ref{fig:mag-dic}.
Comparing this profile with the one previously obtained by means of
Eq.~(\ref{eq:QsuI}) (plotted with a dotted line in the same figure), an overall 
shift towards higher values of polarization is discernible, in addition to 
a slightly different substructure in the line-core, that, however, does not 
provide a more accurate fit to the observed profile. 
The inclusion of dichroism does not modify the width of the profile.
By also including dichroism, if we increase the width of the profile by 
assuming a larger value of the Doppler width, the three-peak structure of the 
profile is found to be lost.

\section{Conclusions}
In terms of the same modeling approximations adopted by \citet{Bel07} to 
reproduce the three-peak $Q/I$ profile observed in the second solar spectrum of 
the Ba~{\sc ii} D$_2$ line, it is impossible to reproduce the similar 
three-peak linear polarization profile exhibited in the same spectrum by 
the Sc~{\sc ii} line at 4247~{\AA}.
This is a surprising result given the similarities between the two lines in the 
intensity spectrum, and that the only stable 
isotope of scandium shows hyperfine structure, the physical aspect which 
was found to be at the origin of the triple peak structure of the Ba~{\sc ii} 
D$_2$ line $Q/I$ profile. 
The possibility of reproducing the polarization signal of this ionized scandium 
line in terms of hyperfine structure was also suggested by the three peaks 
having the same amplitude, a circumstance that appears to be in perfect 
agreement with the absence of scandium isotopes without hyperfine structure.
We are indeed able to obtain a $Q/I$ profile with three peaks of the 
same amplitude, but unfortunately its width is found to be considerably smaller 
than that of the observed profile. 
Moreover, while the red dip can be reproduced quite well, the depth of the 
blue one is much smaller than in the observation. 
The only way to increase the width of the profile is to assume a larger 
Doppler width, but, in this case, the three peak structure is rapidly 
lost. 
We investigated the effect of a microturbulent magnetic field, as well as 
the effect of dichroism, but they both do not seem to be able to explain the 
disagreement between our theoretical result and the observation.
Nevertheless, it is important to point out that the lower level of this line, 
being polarizable irrespective of the hyperfine structure ($J_{\ell}=2$), 
is found to be as polarized as the upper level (depolarizing collisions have 
been neglected in this investigation). 
This circumstance explains why the lower-level Hanle effect is so clearly 
evident in this line.
In conclusion, the physical origin of the observed three-peak $Q/I$ profile 
of this scandium line does not seem to be the mere presence of hyperfine 
structure.
We may speculate that an additional physical mechanism is at work 
(such as, for instance, PRD effects). However, if this turns out to be the 
case, it would be of interest to investigate the influence of this physical 
mechanism on the $Q/I$ profile of the Ba~{\sc ii} D$_2$ line, whose triple peak structure could be interpreted solely in terms of hyperfine structure.

\begin{acknowledgements}
The author wishes to express his gratitude to Egidio Landi Degl'Innocenti and 
Javier Trujillo Bueno for several scientific discussions, and for helpful 
comments and suggestions.
\end{acknowledgements}

\end{document}